\documentclass[12pt]{article}
\usepackage{graphicx}

\newcommand\pubnumber{SNSN-323-63}
\newcommand\pubdate{\today}

\def\napoli{Department of Physics\\
Kansas State University, Manhattan, KS, 66502, USA}
\def\support{
	}

\def\Title#1{\begin{center} {\Large #1 } \end{center}}
\def\Author#1{\begin{center}{ \sc #1} \end{center}}
\def\Address#1{\begin{center}{ \it #1} \end{center}}

\newcommand\pubblock{\rightline{\begin{tabular}{l} \pubnumber\\
         \pubdate  \end{tabular}}}
\newenvironment{Abstract}{\begin{quotation}  }{\end{quotation}}
\newenvironment{Presented}{\begin{quotation} \begin{center} 
             PRESENTED AT\end{center}\bigskip 
      \begin{center}\begin{large}}{\end{large}\end{center} \end{quotation}}
\def\Acknowledgements{\bigskip  \bigskip \begin{center} \begin{large}
             \bf ACKNOWLEDGEMENTS \end{large}\end{center}}

\def\beq{\begin{equation}}
\def\eeq#1{\label{#1}\end{equation}}
\def\eeqn{\end{equation}}

\def\beqa{\begin{eqnarray}}
\def\eeqa#1{\label{#1}\end{eqnarray}}
\def\eeqan{\end{eqnarray}}

\let\bar=\overbar

\def\Dslash{\not{\hbox{\kern-4pt $D$}}}
\def\dslash{\not{\hbox{\kern-2pt $\del$}}}

\def\msb{{\bar{\ssstyle M \kern -1pt S}}}

\begin{document}
\begin{titlepage}
\pubblock

\vfill
\Title{Limits on Fourth Generation Fermions}
\vfill
\Author{ Andrew Ivanov~\footnote{on behalf of the ATLAS and CMS Collaborations}\support}
\Address{\napoli}
\vfill
\begin{Abstract}
We present recent results on searches for the fourth generation quarks performed 
at the CMS and ATLAS experiments at LHC. \end{Abstract}
\vfill
\begin{Presented}
Flavor Physics and CP Violation Conference\\
Buzios, Rio, Brazil,  May 20--24, 2013
\end{Presented}
\vfill
\end{titlepage}
\def\thefootnote{\fnsymbol{footnote}}
\setcounter{footnote}{0}

\section{Introduction}

In the standard model of particle physics fermions are arranged in three generations.
The sequential fourth generation is a simplest extension of the experimentally 
established standard model.
Electroweak precision measurements favor small mass splitting between fourth generation 
quarks $m(t')-m(b') < M_W$~\cite{kribbs, he}. Assuming nearly diagonal structure of the CKM 
4$\times$4 matrix, the fourth generation quarks then preferentially decay as $t' \to Wb$, and $b' \to tW$.

Discovery of the Higgs boson and recent measurements of the Higgs production cross section 
in various channels disfavor the chiral fourth generation, since this model predicts a specific 
hierarchy of signal strengths~\cite{1111.6395}, which are not supported by experimental results~\cite{nierste,cms-higgs,atlas-higgs}, assuming a single Higgs boson. Nevertheless, they might remain viable in the scenarios with the extended Higgs sector~\cite{higgs-doublet, he2}.

More interesting and well-motivated scenarios include an extra family of quarks with different quantum numbers than previous three generations of particles. So called vector-like 
quarks, where both chiralities have the same transformation properties under the electroweak 
symmetry group SU(2)$\times$U(1), appear in Little Higgs models~\cite{little-higgs}, models with warped extra dimensions~\cite{extra-dimensions}, as well as 
in non-minimal supersymmetric extensions~\cite{vector-like-susy}. 
These quarks cancel quadratic divergences in the Higgs mass induced by radiative corrections of
top quark and resolve the hierarchy problem in the particle physics. 
The searches for these quarks represent a very active area of the experimental research at both CMS~\cite{cms} and ATLAS~\cite{atlas} experiments. In this article we present recent results from the searches for extra family of quarks.

\section{Vector-Like Quarks Phenomenology}

Vector-like quarks can be produced at LHC in pairs via the strong interaction or singly via 
the electroweak interaction. Pair production cross sections are computed at approximate NNLO using 
HATHOR (Hadronic Top and Heavy quarks crOss section calculatoR)~\cite{HATHOR}.
The single production depends on the unknown electroweak couplings of the new quarks.
It has much milder dependence on the mass of new quarks, and it is expected to dominate 
the pair production at higher masses.

In most models the vector-like quarks couple to the 3rd generation quarks of the standard model.
The flavor-changing-neutral-current decays are allowed, and at the tree level the vector-like quarks can decay into the following modes:
$t' \to Wb, Zt, Ht$, and $b' \to Wt, Zb, Hb$.
In some models the branching ratios into these modes is fixed, and is a function of the new 
heavy quark mass. If the quark is a weak-isospin singlet, all three decay modes are allowed, while for the weak-isospin doublet one of the decay products must be a neutral Z or Higgs boson~\cite{top_partners}.

\section{Searches for Up-Type Quark}

The decay $t' \to Wb$ is identical to the standard model top quark signature, and 
historically the tools used for the top quark event reconstruction were employed to searches 
for the heavy $t'$ quark as well. For $t'$ masses above 500 GeV, in a significant fraction
of events a W-boson from $t'$ decay is significantly boosted, and is often reconstructed as 
a single jet. 

\begin{figure}[htb]
\centering
\includegraphics[width=3.1in]{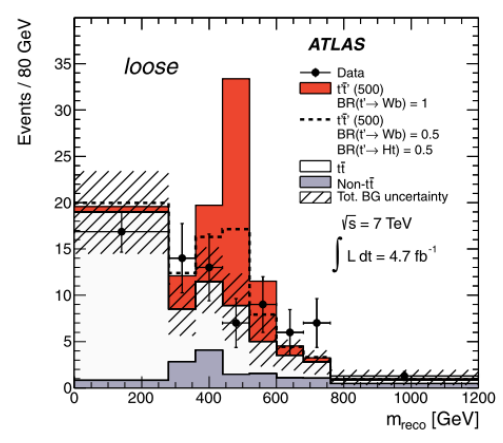}
\includegraphics[width=2.5in]{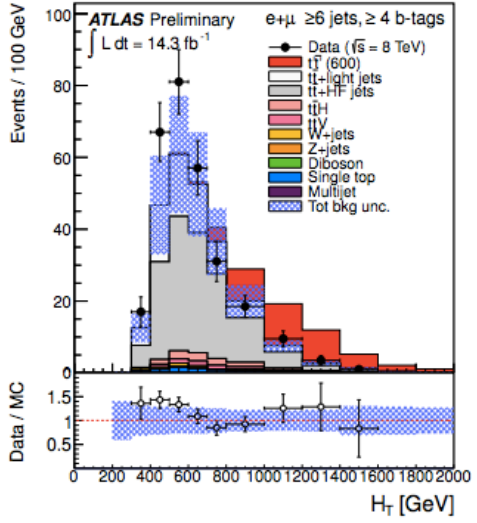}
\caption{
{\bf Left}.
Reconstructed $t'$ mass distribution for $t' \to Wb$ decays. An example of expected signal contribution for $m(t') = 500$ GeV is also shown.
{\bf Right}.
Distribution of $H_T$ in the $\ell + \geq 6$ jets and $\geq 4$ $b$-tags.
The expected $t'\bar{t'}$ signal corresponding to $m(t')=600$ GeV is shown in the $t'$ doublet scenario. }
\label{fig:tprime_Wb}	
\end{figure}

The ATLAS search~\cite{PLB_718_2013_1284} makes use of this fact, and 
consideres two types of hadronically decaying $W$'s: the single jet with transverse momentum 
$p_T > 250$ GeV, and the jet mass consistent with the W boson, or the di-jet pair 
with $\Delta R <$ 0.8 and $p_T > 150$ GeV, and an invariant mass consistent with the W boson.
The search is performed in the lepton plus jets
channel with additional kinematic cuts improving sensitivity to the $t'$ events.
The mass of the heavy $t'$ quark can be reconstructed, as it is shown in Figure~\ref{fig:tprime_Wb} (left). 
The data shows no excess that can be attributed to the new heavy quark, and 
the mass of the $t'$ quark below 656 GeV is excluded, assuming ${\cal B} (t'\to Wb) = 100\%.$
The results of this search can be interpreted as a limit on the branching ratio of ${\cal B} (t'\to Wb)$, 
as a function of the $t'$ mass.  
For $m(t') = 550$ GeV  the branching ratio ${\cal B} (t'\to Wb) > 63\%$ is excluded at 95\% C.L.

The ATLAS search for $t' \to Ht$~\cite{ATLAS-CONF-2013-018}
 is performed using a region orthogonal to the previous search.
Events are classified based on the number of $b$-tags (2,3 or $\geq 4$).
The Figure~\ref{fig:tprime_Wb} (right) shows the scalar transverse energy of all objects 
for events with $\geq 4$ $b$-tags.
The mass of the $t'$ quark below about 850 GeV is excluded, 
assuming ${\cal B} (t'\to Ht) = 100\%.$
The results together with results from the previous search are interpreted as a limit on 
SU(2) doublet and SU(2) singlet quarks, which correspond to 790 and 640 GeV, respectively.

\begin{figure}[htb]
\centering
\includegraphics[width=4.8in]{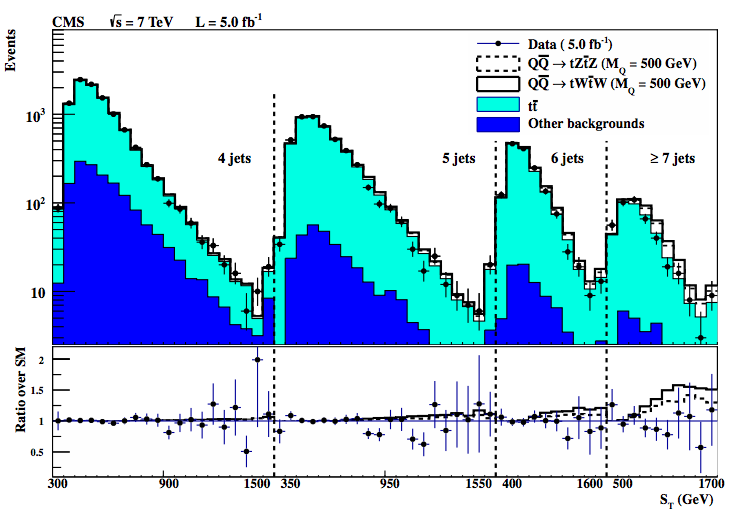}
\caption{Distributions in $S_T$ for different jet multiplicities after the maximum-likelihood fit to data.
The bottom plot shows the ratios of data and background plus signal over the standard model predictions for 
illustrative purposes for the quark mass of 500 GeV.}
\label{fig:tprime_Zt}
\end{figure}

The $t' \to Zt$ events are associated with a high jet multiplicity, and this fact 
is employed in the CMS search~\cite{JHEP_01_2013_154} in the lepton plus jets channel.
The fit to the scalar transverse energy of all objects $S_T$ per event is performed 
as a function of the jet multiplicity (see Figure~\ref{fig:tprime_Zt}),
 excluding $t'$ masses for exclusive $t' \to tZ$ decays 
below 625 GeV at 95\% C.L.
The same results can be used for interpretation of the down-type quark decays $b' \to tW$ 
and correspond to 675 GeV exclusion limits.

\section{Searches for Down-Type Quark}

One can enhance sensitivity to the down-type quark production $b' \to Wt, Zb$ by 
searching for leptonic decays of vector bosons. 
The CMS search~\cite{CMS_PAS_SUS_12_027} requires at least three soft leptons with 
$p_T > 20,10,10$ GeV and at least one $b$-tag, and classifies events based on the number 
of observed leptons, hadronic taus, $b$-jets, and pairs of leptons consistent with $Z$ boson. 
Overall, 240 exclusive channels are exploited. For the channels with sufficient statistics the fit 
to the $S_T$ variable is performed. This analysis sets 760 GeV and 660 GeV limits on mass 
of the $b'$ quark for exclusive $Wt$ and $Zb$ decays, respectively. 
Assuming the $b'$ quark decays into each $Wt$ or $Zb$ mode 50\% of the time, 
the $b'$ mass below 715 GeV is excluded at 95\% C.L.

The ATLAS searched for the $b' \to Wt$ decays using same-sign charge dilepton events~\cite{ATLAS-CONF-2013-051}.
With $9.3\pm0.8$ expected in the signal region, and 15 events observed, the limit is 720 GeV.
For $m(b') = 550$ GeV, the ${\cal B} (b'\to Wt) > 35\%$ is excluded at 95\% C.L.

\section{Search for $T^{5/3}$ Quark}

The quark with an electric charge of $5/3$ appears in the standard model extensions with 
a custodial symmetry in the strong sector including a LR parity~\cite{Contino}.
This quark decays dominantly into $Wt$, making the search similar to the down-type quark searches, with an exception that same-sign charge $W$ bosons originate from the same $T^{5/3}$.

CMS searched for $T ^{5/3} \to Wt$ using same-sign charge di-lepton events~\cite{CMS_PAS_B2G_12_012}, and employing 
the jet substructure techniques to identify hadronically decaying $W$-bosons and top quarks.
In the signal region 6.6 $\pm$ 2.0 events are expected from the standard model backgrounds, 
and 11 events are observed. This corresponds to the observed limit of 770 GeV at 95\% C.L.

\section{Inclusive Search for the Chiral 4th Generation Quark}

\begin{figure}[htb]
\centering
\includegraphics[width=4.8in]{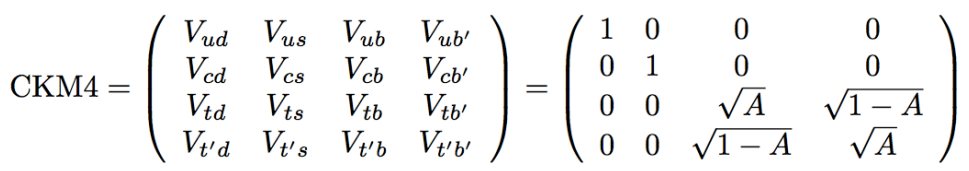}
\caption{The simplified CKM4 matrix parametrization exploited in the inclusive 4th generation search.}
\label{fig:CKM4}
\end{figure}

\begin{figure}[htb]
\centering
\includegraphics[width=2.8in]{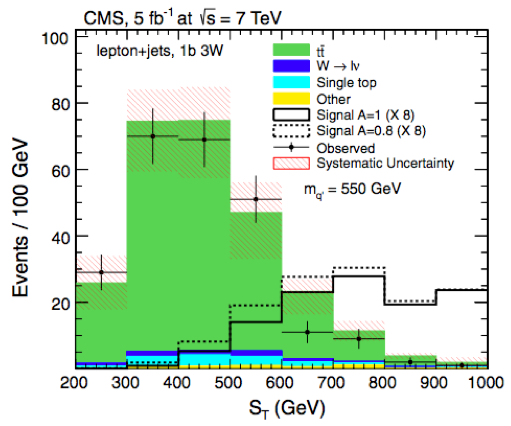}
\includegraphics[width=2.8in]{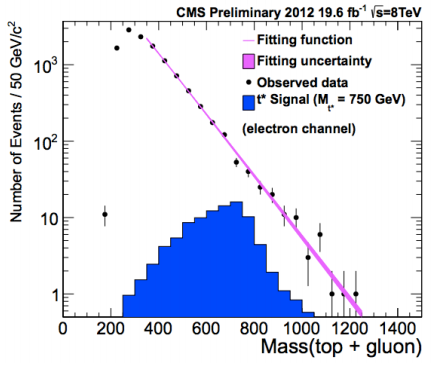}
\caption{{\bf Left:} The $S_T$ distribution for the subsample with one $b$ jet and three $W$ bosons
in the inclusive fourth generation search.
{\bf Right:} The reconstructed top + gluon mass spectrum along with the background fit to the data. 
}
\label{fig:Incl4thGen}
\end{figure}

The CMS performed an inclusive search for the chiral 4th generation in single-lepton, 
same-sign and opposite-sign dilepton and tri-lepton channels simultaneously~\cite{PRD_86_2012_112003}.
For simplicity the CKM4 matrix is parametrized as shown in Figure~\ref{fig:CKM4}, 
using a single parameter $A$, that determines the single electroweak production cross section
of the fourth generation quarks. Events are categorized based on the number of reconstructed $W$ bosons, 
and the fit to the $S_T$ is performed in each channel.
As an example the $S_T$ distribution in the channel with single $b$ and three $W$ bosons
is shown in Figure~\ref{fig:Incl4thGen}.

If $m(t')=m(b')$, the fourth generation quarks with masses below 685 GeV are excluded 
at 95\% C.L. The limit is more stringent if $A < 1$, and the single electroweak production 
is enhanced.

\section{Search for Excited Top Quark $t^* \to tg$}

Some models speculate that the top quark is a composite particle and 
predict an excited top quark with a spin $3/2$ that decays into the standard model 
top and gluon~\cite{excited}.
CMS performed a search for this quark requiring one lepton, 6 and more jets, at least one of which is $b$-tagged~\cite{CMS_PAS_B2G_12_014}.
The mass of the $t^*$ quark is reconstructed using a kinematic $\chi^2$ fitter. 
The backgrounds are modeled using data-driven approach and with a good approximation 
can be fitted using $bg(m) = \frac{a}{1+e^{(m-b)/c}}$ function, as shown in Figure~\ref{fig:Incl4thGen} (right).
This analysis excludes an excited top quark with mass below 794 GeV.

\section{Search for Excited Bottom Quark $b^* \to Wt$}

\begin{figure}[htb]
\centering
\includegraphics[width=5.8in]{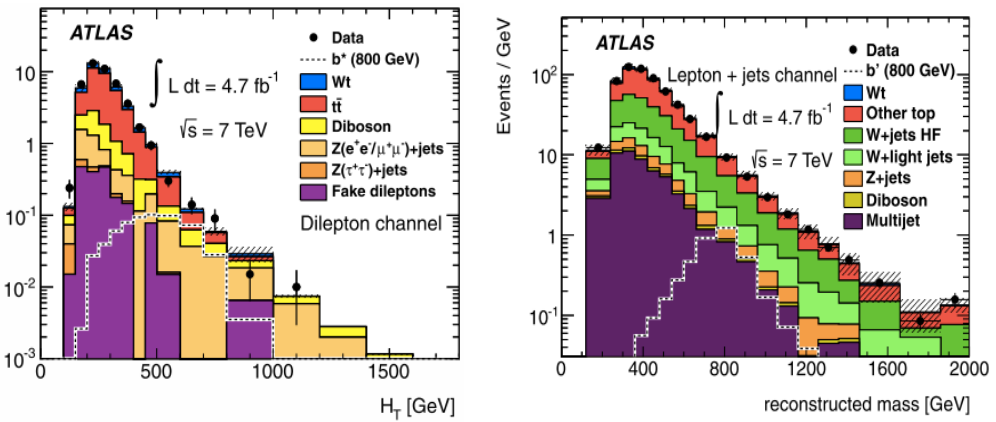}
\caption{{\bf Left:} $H_T$ distribution in the dilepton channel. The hatched band shows 
the uncertainty due to the background normalization. 
{\bf Right:} Reconstructed mass distribution for the lepton + jets channel. The signal for a mass
of 800 GeV is shown in both distributions.
}
\label{fig:single-b}	
\end{figure}

ATLAS performed a search for single $b^*$ quark production that is produced via chromomagnetic interaction, and has an electroweak decay to $Wt$~\cite{PLB_721_2013_171}.
The search is performed simultaneously in the single-lepton and di-lepton channels by fitting 
the data to the $b^*$  reconstructed mass and $H_T$ variables, respectively. 
The data-Monte Carlo agreement is shown in Figure~\ref{fig:single-b}.
For the purely left-handed $b^*$ and a unit strength chromomagnetic coupling the 95\% C.L limit is 870 GeV
on the mass of the $b^*$.

\section{Summary}

The results of searches for the fourth generation quarks are summarized in Table~\ref{tab:sum}.
This list is not exhaustive and more comprehensive searches based on the 20 fb$^{-1}$ dataset at 8 TeV pp collisions 
are underway at both CMS and ATLAS experiments.

\begin{table}[t]
\begin{center}
\begin{tabular}{l|ccc}  
Mass, Model &  95\% C.L limit &  Experiment, Channel \\ \hline
$m(t'), t' \to Wb$  &   $ > 656 $ GeV     &   ATLAS, $\ell + $ jets~\cite{PLB_718_2013_1284} \\
$m(t'), t' \to Ht$  &   $ > 850 $ GeV     &   ATLAS, $\ell + $ jets~\cite{ATLAS-CONF-2013-018} \\
$m(t'), t' \to Zt$  &   $ > 625 $ GeV     &   CMS, $\ell + $ jets~\cite{JHEP_01_2013_154} \\
$m(b'), b' \to Wt$  &   $ > 760 $ GeV     &   CMS, multi-lepton~\cite{CMS_PAS_SUS_12_027} \\
$m(b'), t' \to Zb$  &   $ > 660 $ GeV     &   CMS, multi-lepton~\cite{CMS_PAS_SUS_12_027} \\
Inclusive $t',b'$ & $ > 685 $ GeV     &   CMS, multi-channel~\cite{PRD_86_2012_112003}\\
$m(t')$, SU(2) singlet  &   $ > 640 $ GeV     &   ATLAS, $\ell + $ jets~\cite{ATLAS-CONF-2013-018} \\
$m(b')$, SU(2) singlet  &   $ > 590 $ GeV     &   ATLAS, same-sign dilepton~\cite{ATLAS-CONF-2013-051} \\
$m(t')$, SU(2) singlet  &   $ > 790 $ GeV     &   ATLAS, $\ell + $ jets~\cite{ATLAS-CONF-2013-018} \\
$m(t'), t' \to tg$ &   $ > 794 $ GeV     &   CMS, $\ell + $ jets~\cite{CMS_PAS_B2G_12_014} \\  \hline
\end{tabular}
\caption{Summary of limits on the 4th generation quarks.}
\label{tab:sum}
\end{center}
\end{table}

\Acknowledgements
I would like to thank the organizers for the interesting and enjoyable  conference.

\end{document}